\documentclass[prd,preprint,superscriptaddress,showpacs,nofootinbib,%
tightenlines]{revtex4}
\usepackage{graphics}
\usepackage{epsfig}
\usepackage{color}
\usepackage{slashed}

\newcommand{\ben}{\begin{displaymath}}
\newcommand{\een}{\end{displaymath}}
\newcommand{\be}{\begin{equation}}
\newcommand{\ee}{\end{equation}}
\newcommand{\bea}{\begin{eqnarray}}
\newcommand{\eea}{\end{eqnarray}}
\begin{document}
\title{Electromagnetic transition form factors of the Roper resonance in baryon chiral perturbation theory}
\author{M.~Gelenava}
\affiliation{Center for Elementary Particle Physics, ITP, Ilia State University, \\ 0162 Tbilisi, Georgia
}
\affiliation{Helmholtz Institut f\"ur Strahlen- und Kernphysik and Bethe
  Center for Theoretical Physics, Universit\"at Bonn, D-53115 Bonn, Germany}
\date{Oktober 8, 2017}
\begin{abstract}
We consider the electromagnetic transition form factors of the Roper resonance in the framework of an effective field theory of pions, 
nucleons and delta and Roper resonances as explicit degrees of freedom. Due to the lack experimental data in the region of applicability of low 
energy effective field theory we fit available free coupling constants and compare obtained results compare to the predictions of a theoretical model.

\end{abstract}
\pacs{11.10.Gh,12.39.Fe}
\maketitle

\section{Introduction}
\label{Intro}

   Chiral perturbation 
theory \cite{Weinberg:1979kz,Gasser:1983yg} - low-energy effective field theory (EFT) of QCD
leads to a successful description of the Goldstone boson sector.
It turned out that generalisation of ideas of ChPT to the EFTs with
heavy degrees of freedom is a non-trivial problem.
In baryon chiral perturbation 
theory dimensionally regulated loop diagrams violate the power counting \cite{Gasser:1987rb}. This problem has been solved by using the heavy-baryon 
approach \cite{Jenkins:1990jv,Bernard:1992qa,Bernard:1995dp}
and later by choosing a suitable renormalization scheme
\cite{Tang:1996ca,Becher:1999he,Gegelia:1999gf,Fuchs:2003qc}.
While the $\Delta$ resonance and (axial) vector mesons can also be consistently included  
in low-energy EFT  (see e.g. Refs.~\cite{Hemmert:1997ye,Pascalutsa:2002pi,Bernard:2003xf,Pascalutsa:2006up,Hacker:2005fh,Fuchs:2003sh,Bruns:2004tj,Bruns:2008ub,Terschluesen:2013pya,Leupold:2012qn}), the treatment of heavier baryons such as the Roper resonance is more complicated.  Ref.~\cite{Epelbaum:2015vea} presented new  ideas on the extension of the 
range of applicability of chiral EFT beyond the low-energy region.   

The Roper resonance has been found in a partial wave  analysis of  pion-nucleon 
scattering data long time ago \cite{Roper:1964zza}. 
Since then it has attracted  particular interest because of being the first nucleon resonance 
that exhibits a decay mode into a nucleon and two pions, besides the 
decay into a nucleon and a pion. 
Another interesting feature is that  the Roper appears below
the first negative parity nucleon resonance, the $S_{11}(1535)$. 
Despite of much effort a satisfactory theory of the Roper resonance is still missing.

First steps in direction of addressing  the Roper resonace state in a chiral EFT  
have been made in 
Refs.~\cite{Borasoy:2006fk,Djukanovic:2009gt,Long:2011rt,Bauer:2012at,Epelbaum:2015vea}.
The pole mass and the width of the 
Roper resonance has been calculated at one-loop order in Refs.~\cite{Borasoy:2006fk,Djukanovic:2009gt} and at two-loop order in Ref.~\cite{Gegelia:2016xcw},
the magnetic moment was studied in Ref.~\cite{Bauer:2012at}.  The impact of the explicit inclusion of the Roper 
resonance in chiral EFT on the $P_{11}$ pion-nucleon scattering phase shifts has been studied in Ref.~\cite{Long:2011rt}.
Recently electromagnetic transition form factors of the Roper resonance have been analysed in  and EFT of pions, nucleons, Roper resonance and vector mesons \cite{Bauer:2014cqa}.   

In this work we calculate the electromagnetic transition form factors of the Roper resonance in a systematic expansion 
in the framework of baryon chiral perturbation theory with pions, nucleons, the delta and 
Roper resonances as explicit degrees of freedom.  Unlike the Ref.~\cite{Bauer:2014cqa}  we consider only small values of the virtual photon momentum square $q^2\sim M_\pi^2$.   
 
The paper is organized as follows: in Section~\ref{EffL}  we specify the effective Lagrangian, 
in Section~\ref{sec:FFs} the transition form factors are defined contributing Feynman diagrams are identified.
Section~\ref{numerics} contains the  
numerical results  and we summarize in Section~\ref{summary}.

\medskip

\section{Effective Lagrangian}
\label{EffL}

    Below we specify the terms of the chiral effective Lagrangian which are  relevant 
for the calculation of the electromagnetic transition form factors of the Roper at next-to-leading order. We consider  pions,
nucleons and the delta and Roper resonances as dynamical degrees of freedom.
The effective Lagrangian can be written as
\bea
{\cal L}_{\rm eff}={\cal L}_{\pi\pi}+{\cal L}_{\pi N}+{\cal L}_{\pi \Delta}+{\cal L}_{\pi R}
+{\cal L}_{\pi N\Delta}+{\cal L}_{\pi NR}+{\cal L}_{\pi\Delta R},
\eea
where the subscripts indicate the dynamical fields which contribute to a given term. 
From the purely mesonic sector we only need the leading order terms gives by \cite{Gasser:1983yg,Bellucci:1994eb}
\bea {\cal
L}_{\pi\pi}^{(2)}&=&  \frac{F^2}{4} {\rm Tr} \left[ \partial_\mu U \partial^\mu U^\dagger \right] 
+\frac{F^2 M^2}{4}\left[ U^\dagger+ U\right] +i \, \frac{F^2}{2} \left[ (\partial^\mu U U^\dagger +\partial^\mu U^\dagger U ) v_\mu \right] ,
\eea
where $F$ is the pion decay constant in the chiral 
limit and $M$ is the leading
term in the quark mass expansion of the pion mass 
\cite{Gasser:1983yg}. The external elctromagnetic field ${\cal A}_\mu$ is represented in $v_\mu =-e {\cal A}_\mu \tau_3/2$, where  $e^2/(4\pi) \approx 1/137$ ($e> 0$). 
The pion field in contained in the unimodular unitary $2\times 2$
matrix $U$. 

The terms of the effective Lagrangian  with pions and baryons needed for our calculation are given by:
\bea
{\cal L}_{\pi N}^{(1)}&=&\bar{\Psi}_N\left\{i\slashed{D}-m+\frac{1}{2}g \,\slashed{u}\gamma^5\right\}\Psi_N\, ,\nonumber\\
{\cal L}_{\pi R}^{(1)}&=&\bar{\Psi}_R\left\{i\slashed{D}-m_R+\frac{1}{2}g_R\slashed{u}\gamma^5\right\}\Psi_R\, ,\nonumber\\
{\cal L}_{\pi R}^{(2)}&=&\bar{\Psi}_R\left\{c_1^R\langle\chi^+\rangle\right\}\Psi_R\, ,\nonumber\\
{\cal L}_{\pi NR}^{(1)}&=&\bar{\Psi}_R\left\{\frac{g_{\pi NR}}{2}\gamma^\mu\gamma_5 u_\mu\right\}\Psi_N+ {\rm h.c.}\, ,\nonumber\\
{\cal L}_{\pi NR}^{(2)}&=&\bar{\Psi}_R\left\{\frac{c_6^*}{2}\, f^+_{\mu\nu}
+\frac{c_7^*}{2} \, v_{\mu\nu}^{(s)}   \right\} \sigma^{\mu\nu}\Psi_N+h.c.\nonumber\\
{\cal L}_{\pi NR}^{(3)}&=& \frac{i}{2} \, d^*_6 \, \bar{\Psi}_R  [D^\mu, f_{\mu\nu}^+] D^\nu \Psi_N +2 i\, d^*_7 \, \bar{\Psi}_R  ( \partial^\mu v_{\mu\nu}^{(s)}) D^\nu \Psi_N+ h.c.\nonumber\\
{\cal L}^{(1)}_{\pi\Delta}&=&-\bar{\Psi}_{\mu}^i\xi^{\frac{3}{2}}_{ij}\left\{\left(i\slashed{D}^{jk}-m_\Delta\delta^{jk}\right)g^{\mu\nu}
-i\left(\gamma^\mu D^{\nu,jk}+\gamma^\nu D^{\mu,jk}\right) +i \gamma^\mu\slashed{D}^{jk}\gamma^\nu+m_\Delta\delta^{jk} \gamma^{\mu}\gamma^\nu\right.
\nonumber\\
 &&\left.+\frac{g_1}{2}\slashed{u}^{jk}\gamma_5g^{\mu\nu}+\frac{g_2}{2} (\gamma^\mu u^{\nu,jk}+u^{\nu,jk}\gamma^\mu)\gamma_5+\frac{g_3}{2}\gamma^\mu\slashed{u}^{jk}\gamma_5\gamma^\nu \right\}\xi^{\frac{3}{2}}_{kl}{\Psi}_\nu^l\,  ,\nonumber
 \\
{\cal L}^{(1)}_{\pi N\Delta}&=&h\,\bar{\Psi}_{\mu}^i\xi_{ij}^{\frac{3}{2}}\Theta^{\mu\alpha}(z_1)\ \omega_{\alpha}^j\Psi_N+ {\rm h.c.}\, ,\nonumber\\
{\cal L}^{(1)}_{\pi \Delta R}&=&h_R\,\bar{\Psi}_{\mu}^i\xi_{ij}^{\frac{3}{2}}\Theta^{\mu\alpha}(\tilde{z})\ \omega_{\alpha}^j\Psi_R+ {\rm h.c.}\, .
\label{LagrNRDP}
\eea
Here $\Psi_N$ and $\Psi_R$ are isospin doublet fields of the nucleon and the Roper resonance, 
respectively.  The $\Delta$ resonance is represented by the vector-spinor isovector-isospinor
Rarita-Schwinger field  $\Psi_\nu$  
\cite{Rarita:1941mf}, 
$\xi^{\frac{3}{2}}$ is the isospin-$3/2$ projector, $\omega_\alpha^i=\frac{1}{2}\, {\rm Tr} \left[ \tau^i u_\alpha \right]$ and $\Theta^{\mu\alpha}(z)=g^{\mu\alpha}
+z\gamma^\mu\gamma^\alpha$ with an off-shell parameter $z$. We fix the off-shell structure 
of the interactions involving the delta by taking $g_2=g_3=0$ and $z_1=\tilde{z}=0$. 
It has been shown that these off-shell parameters can be absorbed in LECs \cite{Tang:1996sq,Pascalutsa:2000kd,Krebs:2009bf}.
Further building blocks are given by 
\begin{eqnarray}
u & = & \sqrt{U}\,,\nonumber\\
\chi^+&=&M^2(U^\dag+ U)\,,  
   \nonumber\\
u_\mu & = & i \left[u^\dagger \partial_\mu u -u \partial_\mu u^\dagger
-i\,\left( u^\dagger v_\mu u-u v_\mu u^\dagger\right) \right], 
\nonumber\\
D_\mu \Psi_{N/R} & = & \left( \partial_\mu + \Gamma_\mu  -i\,v_\mu ^{(s)}
\right) \Psi_{N/R}\,, \nonumber\\
\left(D_\mu\Psi\right)_{\nu,i} & = &
\partial_\mu\Psi_{\nu,i}-2\,i\,\epsilon_{ijk}\Gamma_{\mu,k} \Psi_{\nu,j}+\Gamma_\mu\Psi_{\nu,i}
\,,\nonumber\\
\Gamma_\mu & = &
\frac{1}{2}\,\left[u^\dagger \partial_\mu u +u
\partial_\mu u^\dagger -i\,\left( u^\dagger v_\mu u+u v_\mu u^\dagger\right)
\right]=\tau_k\Gamma_{\mu,k}\,,\nonumber\\
v_{\mu\nu}^{(s)} &=& \partial_\mu v_\nu^{(s)} - \partial_\nu v_\mu^{(s)} , \nonumber\\
f_{\mu\nu}^{+} &=& u f_{\mu\nu} u^\dagger +u^\dagger f_{\mu\nu} u \,, \nonumber\\
f_{\mu\nu} &=& \partial_\mu v_\nu - \partial_\nu v_\mu -i [v_\mu,v_\nu]\,. \label{cders}
\end{eqnarray}

\noindent
A mixing kinetic term of the form
$
i\lambda_1\bar{\Psi}_R\gamma_\mu D^\mu\Psi_N-\lambda_2\bar{\Psi}_R\Psi_N+  {\rm h.c}.
$
is not included in the effective Lagrangian since, using field transformations   
it can be reduced to the form specified above \cite{Borasoy:2006fk}.

\section{Electromagnetic transition form factors of the Roper resonance}
\label{sec:FFs}

Most general parametrisation of the $\gamma^*N R$ vertex function contains three independent Lorentz structures. However due to the current conservation the coefficients of these structures satisfy one identity. 
Therefore the renormalized vertex function of the $R\to \gamma^* N$ transition can be parameterized as 
\begin{eqnarray}
&& \sqrt{Z_R} \ \bar\omega^i(p_f) \Gamma^\mu(p_f, p_i) u^j(p_i) \sqrt{Z_N} \nonumber \\ 
&& \ \ \ = \bar\omega^i(p_f) \Biggl[  \left( \gamma^\mu-\slashed q \,\frac{q^\mu}{q^2} \right) \tilde F_1^*(Q^2) 
 + \frac{i\,\sigma^{\mu\nu} q_\nu}{M_R+m_N}  \tilde F_2^*(Q^2)  \Biggr] u^j(p_i) \sqrt{Z_N},
\label{TrFFs}
\end{eqnarray}
where $Q^2=-(p_f-p_i)^2$ and $Z_R$ and $Z_N$ are the residues of the dressed propagators of the Roper and the nucleon, respectively.

\medskip

\begin{figure}[t]
\epsfig{file=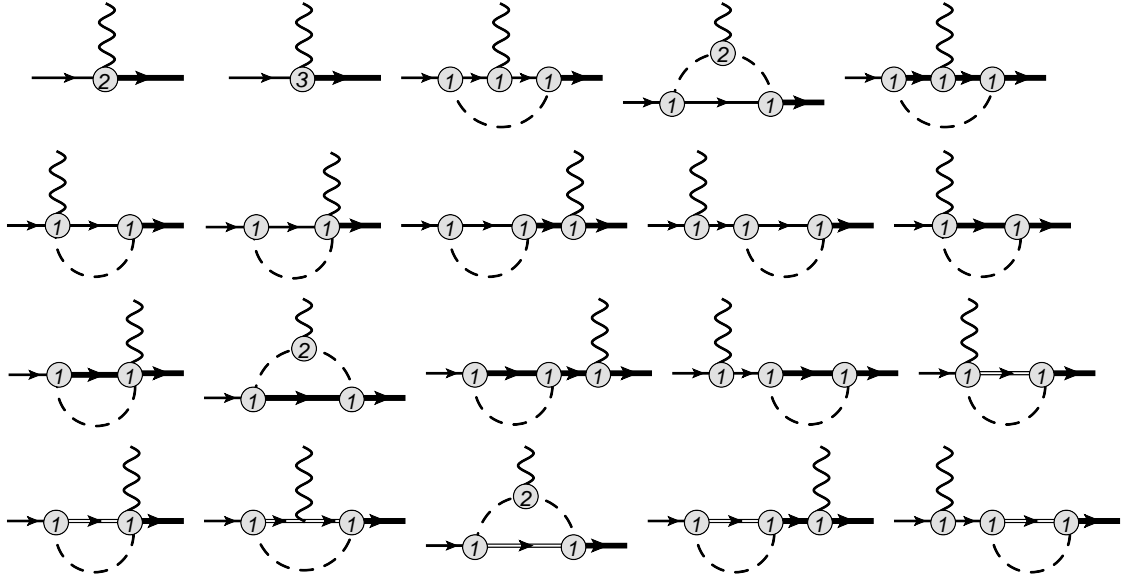,width=0.9\textwidth}
\caption[]{\label{DeltaMassInd:fig} 
Tree and one-loop diagrams contributing to the electromagnetic transition form factors 
of the Roper resonance. The dashed, solid, thick-solid and double-solid lines
correspond to the pion, nucleon, Roper resonance  and delta, respectively. Numbers in vertices indicate the corresponding order of the interaction term.}
\end{figure}

By counting the mass differences $m_R-m_N$, $m_\Delta-m_N$ and $m_R-m_\Delta$ as of the 
same order as the pion mass and the pion momenta, i.e. $\sim q^1$ the standard counting can be applied to all diagrams contributing to the considered transition form factors.
According to the rules of this counting a four-dimensional loop integration is 
of order $q^4$,  an interaction vertex obtained from
an ${\cal O}(q^n)$ Lagrangian counts as of order $q^n$, a pion
propagator as order $q^{-2}$, and a nucleon propagator as order
$q^{-1}$. Further, order $q^{-1}$ is assigned to the
$\Delta$ and the Roper resonance propagators  for non-resonant kinematics. The propagators of 
the delta and the Roper  resonance get enhanced for resonant kinematics when they 
appear as intermediate states outside the loop integration \cite{Pascalutsa:2002pi}. In this case order $q^{-3}$ is assigned to these propagators. 
Up to nex-to-next-to leading order according to the above counting there are tree and one loop diagrams contributing to the electromagnetic transition form factors of the Roper resonance, shown in Fig.~\ref{DeltaMassInd:fig}.
However, due due to the large mass difference $m_R-m_N\sim 400 \, {\rm MeV} \gg M_\pi=135 $ MeV, the above mentioned power counting 
cannot be trusted. Assigning order $\delta^1$ to $m_R-m_N$ it is  more appropriate to count $M_\pi\sim \delta^2$. While different contributions of the same order according the standard counting now become of unequal importance, 
still contributions to the transition form factors up to next-to-next-to leading order are generated by the same diagrams, shown in Fig.~\ref{DeltaMassInd:fig}. Loop diagrams calculated using the standard dimensional regularisation do not satisfy the power counting, however all power counting violating pieces are polynomial in momenta and the squared pion mass and hence can be absorbed in the renormalization of the coupling constants. We have explicitly verified that this is the case.

Notice that loop contributions to the residues of the nucleon and the Roper resonance propagators multiplied with the tree order contributions to the transition form factors start the order higher
than the accuracy of our calculation and therefore we do not take them into account.

\section{Numerical results}
\label{numerics}

To obtain numerical results for the transition form factors we use the following standard 
values of the parameters \cite{Agashe:2014kda} 
\begin{eqnarray}
&& M_\pi = 139 \ {\rm MeV}, \ \ m_N=939 \ {\rm MeV}, \  \  m_\Delta=1210\pm1\ {\rm MeV}, \  \  
\Gamma_\Delta=100\pm2\ {\rm MeV},  \nonumber\\ 
&& \ \  m_R=1365\pm15 \ {\rm MeV},  F_\pi=92.2 \ {\rm MeV} , \ \  g_A=1.27.
\label{Nvalues}
\end{eqnarray}
Further we substitute 
$h = 1.42\pm 0.02$, where the latter is the real part of this coupling taken 
from Ref.~\cite{Yao:2016vbz} (the given error takes into 
account only the statistical uncertainties). The imaginary part of $h$ only contributes
to orders beyond the accuracy of our calculations and therefore we do not include it here. 
We use $g_{\pi NR}=
\pm(0.47\pm0.05)$ 
\cite{Gegelia:2016xcw} 
and in what follows we take both signs into 
account which contributes to the error budget.  
Further, we assume $g_R=g_A$ ($g_A$ is axial coupling of pion and nucleon, difference between $g$ and $g_A$ is in higher order and can be neglected) and $h_{R}=h$, 
which corresponds to the maximal mixing 
assumption of  Ref.~\cite{Beane:2002ud}, where the $\Delta$ and Roper resonances and the nucleon are considered 
as chiral partners in a reducible representation of the full QCD chiral symmetry group.

\begin{figure}[t]
\epsfig{file=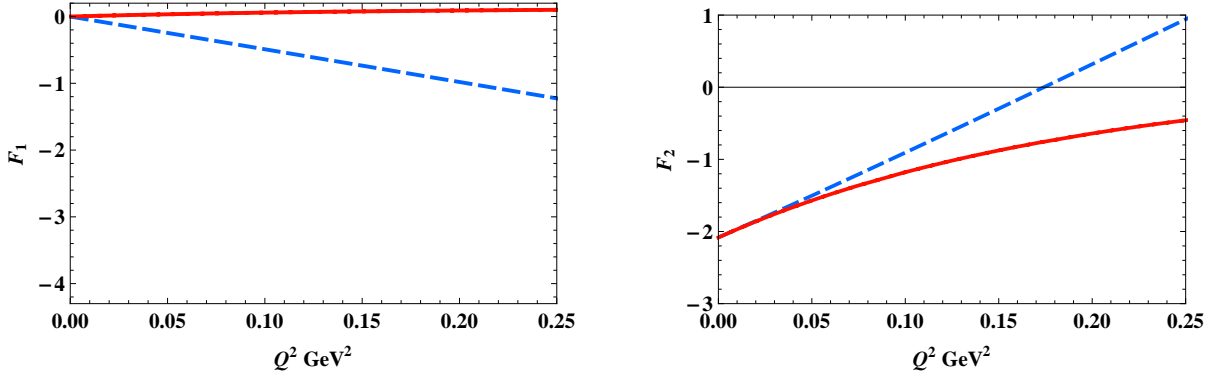,width=1.\textwidth}
\caption[]{\label{NumRes}
Numerical results for the transition form factors.  Solid (red) lines correspond to the parameterisation of  Ref.~\cite{Ramalho:2017muv} and dashed (blue) lines represent the results fo the current work. 
}
\end{figure}

\medskip

As there is no experimental data available for low $q^2$, we used the numerical results obtained from the parameterisation of Ref.~\cite{Ramalho:2017muv}. 
We fixed the unknown parameters contributing at tree order by reproducing the  $F_2^*$ form factor of the proton at low $q^2$.

In Fig.~\ref{NumRes} we plot our  obtained results.

\section{Summary}
\label{summary}

\medskip

   In current work we have  calculated the electromagnetic transition form factors of the Roper resonance up to next-to-leading 
order in a systematic expansion of baryon chiral perturbation theory with pions, nucleons, delta 
and Roper resonances as dynamical degrees of freedom. 

So far experimental data is not available in the region of applicability of chiral effective field theory. Therefore we fitted available free parameters and compared 
our results to predictions of the theoretical parametrization of Ref.~\cite{Ramalho:2017muv}.

\acknowledgments

The author is thankful to J. Gegelia for current interest in the work and useful discussions. This research is  supported in part by Volkswagenstiftung
under contract no. 86260 and by Shota Rustaveli National Science Foundation (SRNSF), grant no. DI-2016-26 "Three-particle problem in a box and in the continuum".

\end{document}